\documentclass[journal=jacsat,manuscript=article]{achemso}
\usepackage{mathpazo}
\usepackage{microtype}
\usepackage{float}
\usepackage{geometry}
\usepackage{natbib}
\usepackage{setspace}
\usepackage{xkeyval}
\usepackage{chemformula}
\usepackage{xcolor}
\makeatletter
\newcommand*{\rom}[1]{\expandafter\@slowromancap\romannumeral #1@} %roomalaiset numerot

\author{Kati Asikainen$^\dagger$}
\email{Kati.Asikainen@oulu.fi}
\author{Matti Alatalo$^\dagger$}
\author{Marko Huttula$^\dagger$}
\author{S. Assa Aravindh$^\dagger$,}
\email{Assa.SasikalaDevi@oulu.fi}
\affiliation[University of Oulu]
{{}$^\dagger$Nano and Molecular Systems Research Unit, University of Oulu, FI-90014, Finland}

\title{Tuning the Electronic Properties of Two-Dimensional Lepidocrocite Titanium Dioxide Based Heterojunctions}

\begin{document}
\clearpage

\begin{abstract}
%Max 150 words
Two-dimensional (2D) heterostructures reveal novel physicochemical phenomena at different length scales, that are highly desirable for technological applications. We present a comprehensive density functional theory study of van der Waals (vdW) heterostructures constructed by stacking 2D \ch{TiO2} and 2D MoSSe monolayers to form \ch{TiO2}-MoSSe heterojunction. The heterostructure formation is found to be exothermic, indicating stability. We find that by varying the atomic species at the interfaces the electronic structure can be considerably altered, due to the differences in charge transfer, arising from the inherent electronegativity of the atoms. We demonstrate that the heterostructures possess a type \rom{2} or type \rom{3} band alignment, depending on the atomic termination of MoSSe at the interface. The observed charge transfer occurs from MoSSe to \ch{TiO2}. Our results suggest that Janus interface enables the tuning of electronic properties, providing understanding of the possible applications of the \ch{TiO2}-MoSSe heterostructure.
\end{abstract}

%Required!!!
%\begin{tocentry}

%\end{tocentry}
%Required!!!
%\begin{tocentry}
%check the size!
%\end{tocentry}

\clearpage   

\section{Introduction}
Titanium dioxide (\ch{TiO2}) is one of the premier materials in various applications including e.g. photovoltaic cells \cite{PV1, PV3}, photocatalysis \cite{PC1, PC2} and batteries \cite{B1,B2}. Excellent chemical stability, eco-friendliness, and low cost are the favorable factors of \ch{TiO2}, but possible drawbacks that limit the performance include large band gap of bulk phases and fast recombination of electron and hole pairs. Various strategies such as doping, tuning of morphology, and constructing heterostructures with different lattice matching semiconductors have been successfully employed in order to overcome the above shortcomings. Another strategy to enhance the activity of semiconductors is to construct low dimensional materials that provide many active surface sites compared to their bulk counterparts. Because of this advantage, two-dimensional (2D) materials are gathering wide attention compared to their 3D phases \cite{2D}. 2D \ch{TiO2} with lepidocrocite-like structure has been synthesized through exfoliation by means of soft-chemical procedures by Sasaki et.al. \cite{TiO21996}, and later theoretically confirmed to be thermodynamically stable \cite{LNS-structure}. Experimentally synthesized 2D Lepidocrocite-type \ch{TiO2} possesses a large band gap (3.8 eV) due to the quantum confinement \cite{LNS-bg}. Despite this shortcoming, it has been shown to be a suitable candidate for both hydrogen and oxygen evolution reactions (HER and OER), with the possibility of improving the photocatalytic performance via transition-metal doping \cite{TiO2-OER, TiO2-HER}. 

For application purposes, heterojunctions constructed by stacking two or more 2D monolayers are promising, as they provide ample opportunities for band bending, due to the spatial variation of the Fermi level of the semiconductors constituting the heterojunction. The three most conventional band alignments are type \rom{1} (straddling gap), type \rom{2} (staggered gap) and type \rom{3} (broken gap), each of these are beneficial for the development of materials for different applications. Type \rom{1} band alignment is useful in optical devices such as light-emitting diodes (LEDs) \cite{LED} as it leads to charge carrier accumulation in one location and high recombination rate under light irradiation. By providing spatial separation of electrons and holes into different locations, thus reducing the recombination rate, type \rom{2} band alignment is desirable in photocatalytic applications \cite{catalysis} and photovoltaic cells \cite{PC}. Finally, the type \rom{3} band alignment allows tunneling of electrons from one material to another, becoming favorable for tunnel field-effect transistors (TFETs) \cite{TFET}. Multiple studies have demonstrated vdW heterostructures and the formation of a heterojunction to improve the light harvesting into visible light, and enable charge transfer across the interface \cite{C3N/MX,MoSSe-WF2,In2Se3/MoS2,XSSe/SiC, MoSSe-GaAs}, leading to superior properties, and broadening the applications of 2D materials.

2D Janus materials are a novel class of  materials, extensively studied recently due to the multitude of opportunities in device applications. They were experimentally synthesized for the first time in 2017, by breaking the out of plane structural symmetry of \ch{MoS2}, and replacing S atoms by Se atoms on one side \cite{Janus2017}. The name Janus originates from the two faced  Roman god Janus, and the MoSSe Janus material consists of two different chalcogen atoms (S and Se) on either side of a Mo atom sandwiched in the middle. Thermodynamic stability of MoSSe is well established from phonon band structure analysis \cite{MoSSe}, and therefore it is worthwhile to investigate if MoSSe can form heterojunctions with other lattice matching semiconductors. Even though the synthesis of MoSSe kick-started the research interest in these materials, recent studies also focus on other possible materials, such as PtSSe, WSeTe and many others, \cite{PtSSe, WSeTe, WSSe, JL} for various applications. 

%[ need to elaborate here why we need heterojunctions, not only monolayers, then the story goes to say why we made TiO2/MoSSe heterojunction.]
%Afterwards many theoretical studies were conducted in focusing on their applications in water splitting, as they kkkt

Previously, 2D lepidocrocite-type \ch{TiO2}-based vdW heterostructures containing GaSe and \ch{MoS2} have been investigated for enhancing the performance of the isolated \ch{TiO2} monolayer \cite{TiO2-GaSe, TiO2/MoS2}.  However, to the best of our knowledge, a vdW heterostructure of 2D \ch{TiO2} and 2D MoSSe has not yet been studied. These materials have fairly similar lattice parameters ($a=3.00$\ \AA\ and $b=3.80$\ \AA\ for 2D \ch{TiO2} \cite{lc-TiO2}, and $a=b=3.24$\ \AA\ for 2D MoSSe \cite{Janus2017}) which is essential in creating small lattice-mismatch heterostructures. Strict lattice-matching may not be necessary \cite{mismatch} but a large mismatch can affect the stability and performance of the heterostructures. Therefore, in this work, we investigate the structural and electronic properties of 2D \ch{TiO2}/MoSSe vdW heterostructure by employing first principles calculations. The study was started by optimizing the heterostructures. The electronic structure was examined through the band structure and density of states and further, charge density analysis, planar-averaged electrostatic potential, and work function were calculated to obtain more insight into charge transport properties in the heterostructures. %The results show that different interface terminations of 2D Janus material at the interface modulates the properties of the heterostructure, leading to different band alignments. Our findings suggest that vdW heterostructures of \ch{TiO2} and MoSSe are potential candidate for applications.

\section{Computational methodology}
We performed first principles density functional theory (DFT) calculations to investigate vdW heterostructure of \ch{TiO2} and MoSSe  using the plane wave code VASP (Vienna Ab initio Simulation Package) \cite{vasp1, vasp2,vasp3,vasp4}. The projected augmented wave (PAW) based pseudopotentials with plane wave basis sets were employed \cite{PAW}. A  kinetic energy cutoff of 520\ eV was employed to include the plane waves in the basis set. The exchange-correlation potential was described by  generalized gradient approximation (GGA) in the Perdew-Burke-Ernzerhof (PBE) scheme \cite{GGA}. Since GGA is inadequate to describe the on-site Coulomb interaction between localized d and f electrons, we applied DFT+U to treat the localized Ti 3d electrons in order to obtain more realistic electronic properties. We added a correction of U$_{\mathrm{eff}}=4.5\ eV$ ($U=4.5$, $J=0$)\cite{U} according to the scheme of  Dudarev \textit{et al.} \cite{GGA+U}. Van der Waals interactions between \ch{TiO2} and \ch{MoSSe} were included with the DFT-D2 method of Tkatchenko and Scheffler \cite{vdw}. The  Brillouin zone was sampled according to the Monkhost-Pack scheme \cite{MP} and Gaussian smearing with a width of 0.05\ eV was used. The convergence threshold for energy and forces were set to  $10^{-6}$\ eV and 0.001\ eV \AA$^{-1}$, respectively. We used VESTA \cite{VESTA} for visualization and VASPKIT \cite{vaspkit} for post-processing the outputs of the DFT-calculations.

The initial structure of 2D \ch{TiO2} was constructed according to the structural parameters reported in Ref. \cite{LNS-structure}. 2D MoSSe unit cell was created from the hexagonal unit cell of \ch{MoS2} by replacing one S interface by Se. To sample the 1st Brillouin zone, k-point meshes of $6 \times 6\times 1$ and $5\times 5\times 1$ were used for \ch{TiO2} and MoSSe monolayers, respectively. The vdW heterostructures were constructed via stacking pristine \ch{TiO2} and MoSSe monolayers, with a rectangular supercell of sizes  1x3x1 and 1x2x1, respectively, along the vertical direction. The size of the rectangular unit cell of MoSSe was $a=3.25$\ \AA \ and $b=5.64$\ \AA \ (Figure S1). Due to the lattice mismatch, the constructed heterostructure forms a Moir$\mathrm{\Acute{e}}$ pattern \cite{Moire}. Resulting from this, the stacking configuration is not the same in all regions but in a long range the periodicity appears. Li \textit{et al.} have investigated a few stacking configurations of 2D lepidocrocite-type \ch{TiO2} and 2D \ch{MoS2}. They found that the Moir$\mathrm{\Acute{e}}$ pattern, in which the zigzag direction of \ch{MoS2} and the in-plane edge of \ch{TiO2} with a smaller lattice parameter were aligned in the same direction, was the most stable according to adsorption energies calculated with respect to the interlayer distance \cite{TiO2/MoS2}. Therefore, in this work we focused on this particular stacking configuration of 2D \ch{TiO2} and 2D MoSSe (Figure S2). The lattice mismatch in the x- and y-directions was calculated as $\frac{(a-b)}{a}\times 100\%$, where $a$ and $b$ denote the lattice parameter of \ch{TiO2} and MoSSe monolayers, respectively. This resulted in a lattice mismatch of 7.17\ \% in the x-direction and -0.08\ \% in the y-direction. In the y-direction, the effect of strain is negligible. A vacuum with a thickness of around 23\ \AA\ was added along z-direction to both interfaces to avoid correlation between periodic images. Because Mo-layer is sandwiched by two distinct chalcogen layers, S-layer and Se-layer, heterostructures with two different interfaces can be constructed: \ch{TiO2}-MoSSe (S atoms at the interface) and \ch{TiO2}-MoSeS (Se atoms at the interface). A k-point sampling of $18 \times 5 \times 1$ within the Monkhorst-Pack scheme was adopted. 

\section{Results and discussion}
Before building the heterostructures, we investigated a freestanding 2D \ch{TiO2} and MoSSe monolayers. The optimized geometries of the monolayers are shown in Figure S3. The optimized lattice parameters of \ch{TiO2} were $a=3.03$\AA\ and $b=3.77$\ \AA, and Ti-O distances were 1.85-2.22\ \AA\ (Figure S4). For MoSSe, we found lattice parameters of $a=b=3.25$\ \AA. Mo-S, Mo-Se, Se-Se and S-S distances were 2.42\ \AA, 2.54\ \AA, 3.26\ \AA\ and 3.26\ \AA, respectively. Our calculated lattice parameters are in agreement with existing research \cite{TiO2-GaSe, MoSSe-bg1}. Furthermore, we calculated the electronic band structure of the monolayers (Figure S5). Using the GGA functional, we found a direct band gap of 2.76\ eV at $\Gamma$ for \ch{TiO2} \cite{LNS-structure, LNS-bg2}. Applying the Hubbard correction the band gap of 3.30\ eV was obtained which compares better with the experimental value of 3.8\ eV \cite{LNS-bg} and previously obtained value using the GGA+U \cite{TiO2-GaSe}. Previously, higher-level approximations have also been applied to calculate the electronic structure. Using the HSE06 functional Li \textit{et al.} \cite{TiO2/MoS2} obtained a band gap of 3.87 eV which is extremely close to the experimental value. Besides, Wang \textit{et al.} \cite{GW} have obtained a band gap of 5.97 eV using the GW approximation, and Zhou \textit{et al.} \cite{LNS-bg2} have calculated the band gap using the G$_0$W$_0$+BSE, and reported a band gap of 5.3 eV. Therefore, it can be seen that we need to be careful while comparing different approaches, as over and under-estimation of band gaps can be seen across different functionals. Calculated direct band gap of 1.59\ eV for MoSSe is closer to the experimental band gap of 1.68\ eV \cite{Janus2017} and reported results using the GGA functional \cite{MoSSe-bg1, MoSSe-bg2}. 

\begin{figure}[h!]\centering
\includegraphics[width=1\linewidth]{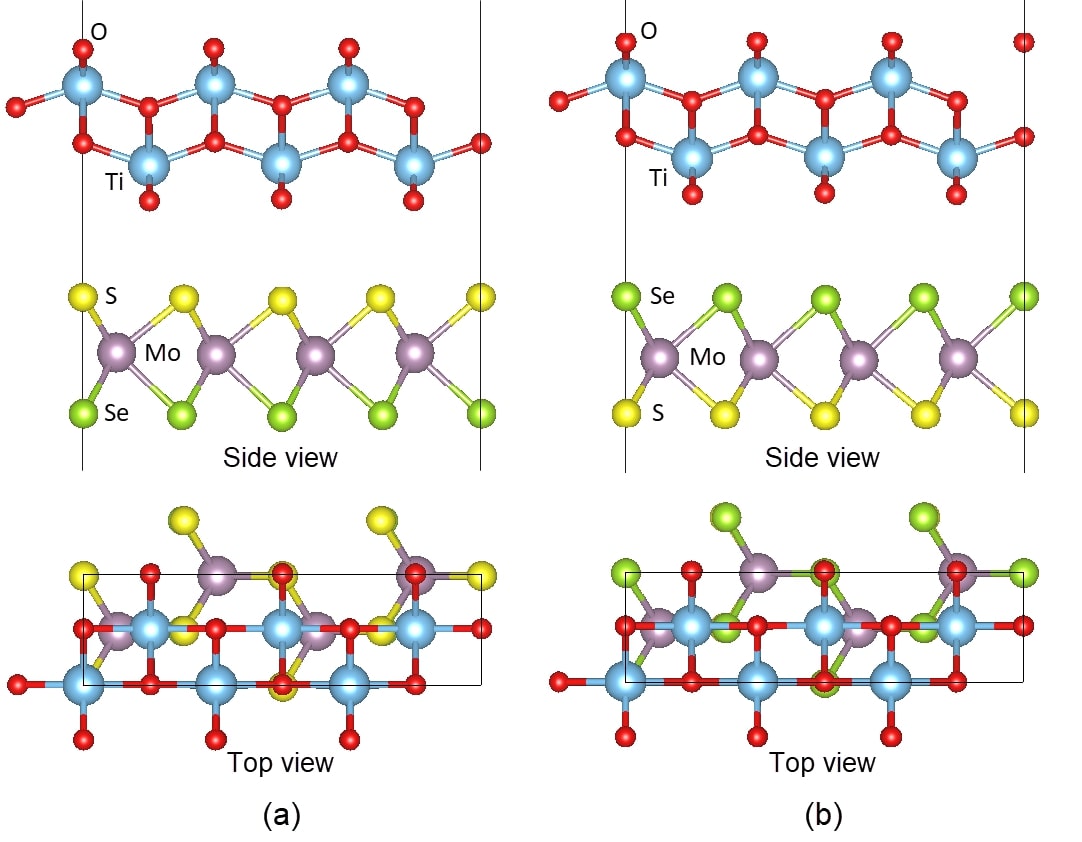}
\caption{Side view (along y-direction) and top view (along z-direction) of the optimized a) \ch{TiO2}/MoSSe and b) \ch{TiO2}/MoSeS heterostructures. The colour coding of the atoms is the same here and elsewhere.}
\label{fig:HSs}
\end{figure}

The optimized structures of the \ch{TiO2}/MoSSe and \ch{TiO2}/MoSeS heterostructures are shown in Figure \ref{fig:HSs}. The obtained lattice constants of the above two heterostructures were $a=3.11$\ \AA\ and  $b=11.18$\ \AA\ after the optimization. The interlayer distance between the monolayers was 2.75\ \AA\ in the \ch{TiO2}/MoSSe and 2.90\ \AA\ in the \ch{TiO2}/MoSeS. These values fall within the optimal range of vdW interaction as discussed by Wang \textit{et al.} \cite{Wang} and Pushkarev \textit{et al.} \cite{H4}. The smaller interlayer distance in the \ch{TiO2}/MoSSe may be attributed to a larger covalent radius of the Se atoms than the S atoms, resulting in a larger spacing between the monolayers \cite{d1, d2}. Interlayer distances smaller than 3 \AA\ have also been reported in previous investigations on \ch{TiO2}-based heterostructures \cite{TiO2-GaSe, TiO2/MoS2, 3D-TiO2-2D-MoS2} and other vdW heterostructures as well \cite{H1, H2, H3}. To estimate the stability of the heterostructures we calculated the formation energies using the equation
\begin{align}
E_F=E_{\mathrm{Heterostructure}}-E_{\ch{TiO2}}-E_{\mathrm{MoSSe}},
\end{align}
where $E_{\mathrm{Heterostructure}}$ is the total energy of the heterostructure and $E_{\ch{TiO2}}$ and $E_{\mathrm{MoSSe}}$ are the total energies of the \ch{TiO2} and MoSSe monolayers. The calculated formation energies of -5.52\ eV and -5.50\ eV for \ch{TiO2}/MoSSe and \ch{TiO2}/MoSeS, respectively, indicated that the vdW heterostructures are energetically favorable. Previously, Ahmad et al. \cite{ahmad} have reported a low binding energy of -5.97 eV for InSe/\ch{PdSe2} heterostructure. Together with the interlayer distances the results suggest high stability for the heterostructure and stronger physical interaction between the monolayers \cite{d1, H2, H3, H4}. Of the two heterostructures the \ch{TiO2}/MoSSe is slightly more stable than the \ch{TiO2}/MoSeS. 

%%%%%%%%%
\begin{figure}[h!]\centering
\includegraphics[width=1\linewidth]{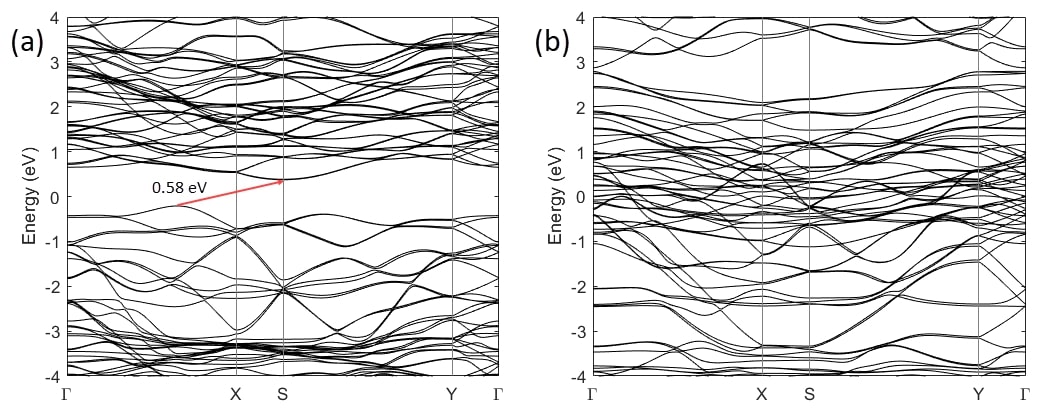}
\includegraphics[width=1\linewidth]{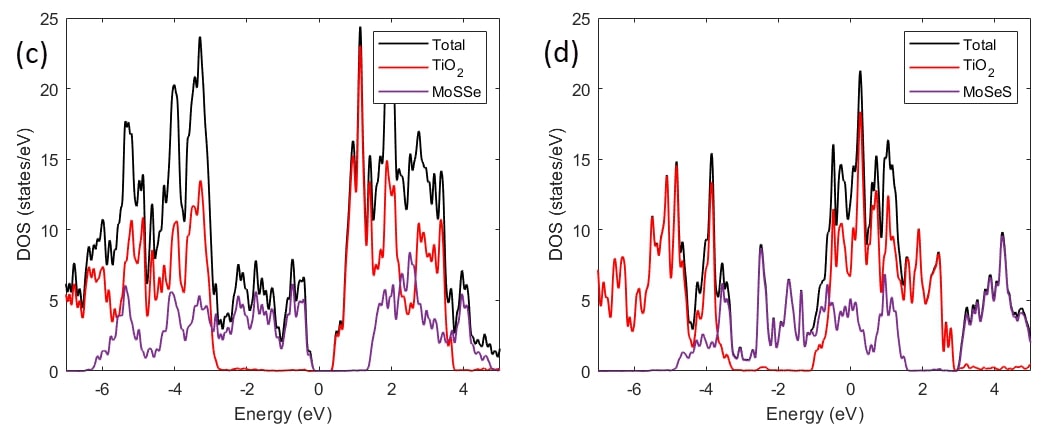}
\caption{Band structure and density of states of the \ch{TiO2}/MoSSe (a and c) and \ch{TiO2}/MoSeS (b and d) heterostructures using the GGA+U functional. The indirect band gap is indicated with a red arrow in a).}
\label{fig:BS-DOS}
\end{figure}
After investigating the electronic structure of the monolayers, we extended the same calculations for the heterostructures. Figure \ref{fig:BS-DOS} shows the band structure and density of states (DOS) of the heterostructures. The formation of the vdW heterostructure led to a significant left-shift of the density of states of \ch{TiO2}, towards the lower energy regions. The \ch{TiO2}/MoSSe heterostructure was found to be an indirect-band gap semiconductor (Figure \ref{fig:BS-DOS}a). The valence band maximum (VBM) was located between $\Gamma$ and X and the conduction band minimum (CBM) at S, resulting in a band gap of 0.58 eV. This is evidently lower than the band gap of the freestanding monolayers, facilitating electron excitation. The quasi-direct band gap was located between S and Y and it was 0.84 eV. It can be seen from the DOS that the top of the valence band (VB) is dominated by the MoSSe monolayer while the bottom of the conduction band (CB) is contributed by \ch{TiO2} (Figure \ref{fig:BS-DOS}c). This indicated a type \rom{2} band alignment between \ch{TiO2} and MoSSe, allowing the charge transfer from MoSSe to \ch{TiO2}. To confirm this, we have calculated the decomposed charge densities of the VBM and CBM which showed that the VBM is constituted from the states of MoSSe while the CBM is contributed by \ch{TiO2} (Figure S6a and S6b). Consequently, the charge density is completely separated, making it possible to separate the photogenerated electrons and holes in the heterostructure. This particular band alignment is extremely desired for photocatalytic applications \cite{catalysis, XSSe/SiC}. The band structure revealed the formation of  metallic states in the \ch{TiO2}/MoSeS due to the overlapping of VB of MoSSe with the CB of \ch{TiO2} (Figure \ref{fig:BS-DOS}b). Both the VBM and CBM of \ch{TiO2} lied below the Fermi level, and the VBM and CBM of MoSSe above the Fermi level (Figure \ref{fig:BS-DOS}d), that is, the highest VB edge of MoSSe was located in higher energy level than the lowest CB edge of \ch{TiO2} (Figure\ref{fig:BS-DOS}d). This suggested that the \ch{TiO2}/MoSeS heterostructure possesses a broken gap type-\rom{3} band alignment. The broken gap enables a band-to-band tunneling (BTBT) mechanism between \ch{TiO2} and MoSSe \cite{C3N/MX}. Thus, electrons in the VB of MoSSe can directly tunnel to the CB of \ch{TiO2} without light absorption or emission. The decomposed charge density plots in Figure S6c and S6d showed that the VBM is contributed by MoSSe while the CBM is distributed around both monolayers. In \ch{TiO2} the CBM mainly concentrated on the atoms at the interface in MoSSe, the charge density distributed around all atoms of MoSSe. Electrons from VBM of MoSSe can be excited by photons to the CBM of MoSSe. Simultaneously, the band alignment allows electron tunneling from the VBM of MoSSe to the CBM of \ch{TiO2}, explaining the observed decomposed charge density distribution. We found a large tunneling window (energy difference between the VB of \ch{MoSSe} and CB of \ch{TiO2}) of around 2.78 eV, indicating greater tunneling probability for electrons. Due to the formation of type \rom{3} band alignment associated with metallic character, the \ch{TiO2}/MoSeS heterostructure can be a potential candidate for tunneling devices, such as TFETs and Esaki diodes \cite{TFET, Esaki}. The particular band alignment can induce a negative differential resistance in heterostructures which is favorable especially in TFETs \cite{NDR, NDR1}. The contributions of each element to the density of states are shown in Figure S7.

\begin{figure}[h!]\centering
\includegraphics[width=1\linewidth]{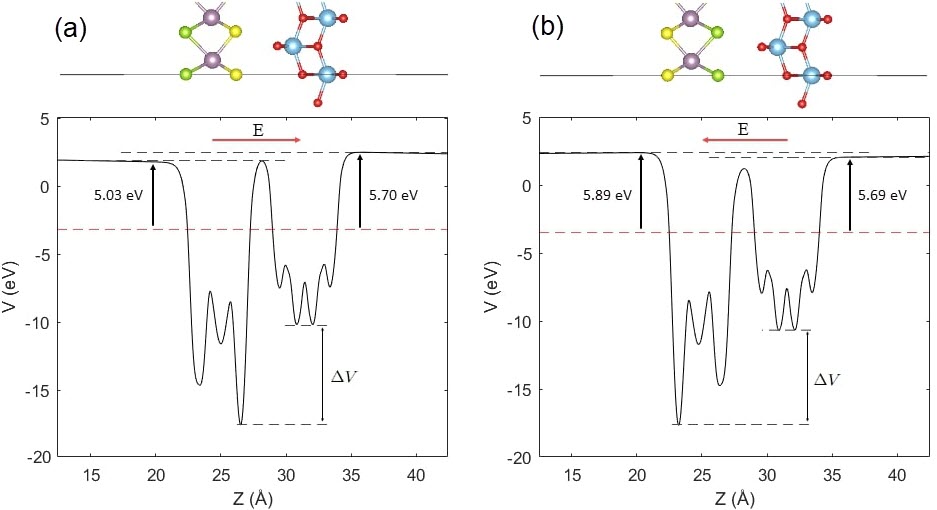}
\caption{Electrostatic potential of the a) \ch{TiO2}/MoSSe and b)\ch{TiO2}/MoSeS heterostructures. The Fermi level is indicated with dashed red line, and the direction of built-in electric field $E$ with red arrow. The potential drop of the heterostructures across the interface is represented by $\Delta V$.}
\label{fig:WFs}
\end{figure}
We calculated the work function $\Phi$ using the equation
\begin{align}
\Phi=E_{\mathrm{vac}}-E_{\mathrm{F}},    
\end{align}
where $E_{\mathrm{vac}}$ and $E_{\mathrm{F}}$ represent the vacuum level and Fermi level, respectively. The calculated work function of \ch{TiO2} was 8.60\ eV (Figure S8a) whereas MoSSe exhibited two different work functions: 5.20\ eV at the Se termination and 5.79\ eV at the S termination, due to the intrinsic polarization (Figure S8b). This resulted in an electrostatic potential difference of 0.59\ eV between the terminations. These are in line with previous work \cite{MoSSe-WF1, MoSSe-WF2}. Because the work function of both the S and Se terminations of MoSSe are smaller than that of \ch{TiO2}, the results suggest electron flow from MoSSe to \ch{TiO2} when combining the two monolayers until thermodynamic equilibrium is reached. We found that the formation of the heterostructures significantly decreased the work function of \ch{TiO2}. Moreover, because of the electrostatic potential difference between the terminations of MoSSe, the work functions of the two surfaces of \ch{TiO2}/MoSSe and \ch{TiO2}/MoSeS were found to be different. The planar-averaged electrostatic potential of the heterostructures along the z-direction is shown in Figure \ref{fig:WFs}. The work functions at the \ch{TiO2} and MoSSe surfaces were 5.70\ eV and 5.03\ eV in the \ch{TiO2}/MoSSe (Figure \ref{fig:WFs}a), and 5.69\ eV and 5.89\ eV in the \ch{TiO2}/MoSeS (Figure \ref{fig:WFs}b), respectively. Interestingly, the heterostructure possesses a lower work function at the MoSSe surface when the S termination is placed at the interface, whereas placing the Se termination at the interface results in a lower work function at the \ch{TiO2} surface. This has also been observed in other 2D MoSSe-based heterostructures \cite{M1, M2, M3}. This may be explained by the intrinsic polarization observed in pristine MoSSe. Since S atoms have larger electronegativity, electrons tend to accumulate in the S layer of MoSSe, increasing the work function and potential energy (Figure S8b). Thus, the direction of the intrinsic dipole moment is from S to Se. When combining MoSSe with \ch{TiO2}, this property appears to be preserved, showing that the intrinsic polarization of MoSSe takes up a significant role in giving rise to a polarization in the heterostructures. Resulting from the different work functions at the two surfaces, there exists an electrostatic potential difference $\Delta \phi$ of 0.67 and 0.2\ eV in the \ch{TiO2}/MoSSe and \ch{TiO2}/MoSeS, respectively, which induces a built-in electric field at the interface of the heterostructures \cite{EF1, MoSSe-WF2, In2Se3/MoS2}, pointing from MoSSe to \ch{TiO2} in the \ch{TiO2}/MoSSe, and from \ch{TiO2} to MoSSe in the \ch{TiO2}/MoSeS. %The lower energy barrier in the \ch{TiO2}/MoSeS can make it easier for electrons to be transferred across the interface. 
Moreover, the electrostatic potential of \ch{TiO2} was deeper than that of MoSSe, resulting in a potential drop $\Delta V$ across the interface. The potential drop was 7.37\ eV in the \ch{TiO2}/MoSSe and 6.95\ eV in the \ch{TiO2}/MoSeS. This gradient, directed from MoSSe to \ch{TiO2}, was attributed to the difference in the electronegativity of oxygen (3.44), and S (2.58) and Se (2.55), and it can further facilitate the charge separation of electrons and holes \cite{Vdrop1, Vdrop2}. 

\begin{figure}[h!]\centering
\includegraphics[width=0.8\linewidth]{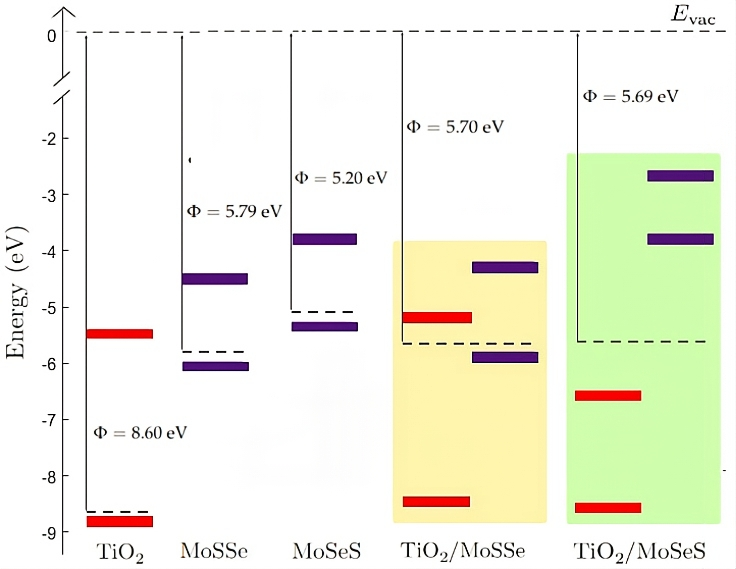}
\caption{Band alignment and work function of the freestanding \ch{TiO2} and MoSSe monolayers, and \ch{TiO2}/MoSSe and \ch{TiO2}/MoSeS heterostructures. The band positions of \ch{TiO2} are depicted in red and MoSSe in purple. The band positions of MoSSe are represented relative to vacuum level by considering the work function of both the S-termination (MoSSe) and Se-termination (MoSeS). The band positions of the \ch{TiO2}/MoSSe and \ch{TiO2}/MoSeS are represented using the work function of \ch{TiO2} (Figure \ref{fig:WFs}). The Fermi level is indicated with a black dashed line, and the vacuum level is set to zero.}
\label{fig:BandDiagram}
\end{figure}

A schematic diagram in Figure \ref{fig:BandDiagram} shows the work functions and band edge positions of the monolayers and the heterostructures with respect to the vacuum level. The band alignment of materials is essential in designing materials for practical applications. The \ch{TiO2}/MoSSe heterostructure retained the band ordering of the two monolayers, resulting in a type \rom{2} band alignment, and the band gap energies of the monolayers were only slightly affected by the formation of the heterostructure. We found a band gap of 3.26 eV for \ch{TiO2} and 1.58 eV for MoSSe. In the \ch{TiO2}/MoSeS the band gap of \ch{TiO2} and MoSSe did not overlap which is known as broken gap. The energy difference between the VB and CB of \ch{TiO2} and the VB and CB of MoSSe were reduced to 2.43 eV and 1.44 eV, respectively. This shows that Se termination at the interface affects more significantly the band gap energies of the monolayers in the heterostructure. 

\begin{figure}[h!]\centering
\includegraphics[width=0.9\linewidth]{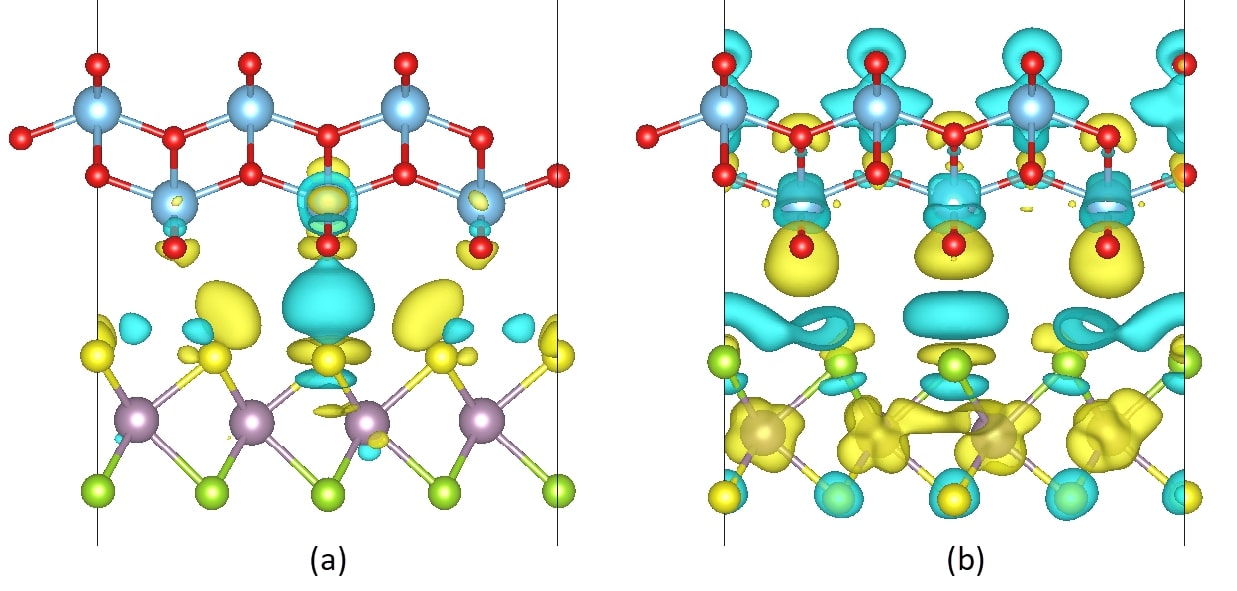}
\caption{Charge density difference of the a) \ch{TiO2}/MoSSe and b) \ch{TiO2}/MoSeS. The yellow isosurface refers to electron gain and blue refers to electron loss. The isosurface value is set to 0.003\ \textit{e\AA$^{-3}$}.}
\label{fig:CDD}
\end{figure}

To further identify the charge transfer in the heterostructures, we have calculated the charge density difference in Figure \ref{fig:CDD} as $\Delta \rho =\rho_{\mathrm{Heterostructure}}-\rho_{\mathrm{\ch{TiO2}}}-\rho_{\mathrm{MoSSe}}$ where $\rho_{\mathrm{Heterostructure}}$, $\rho_{\mathrm{\ch{TiO2}}}$ and $\rho_{\mathrm{MoSSe}}$ are the charge densities of the heterostructure, \ch{TiO2} monolayer and MoSSe monolayer, respectively. In the \ch{TiO2}/MoSSe, the charge redistribution was localized at the interface. In the \ch{TiO2} the O atoms at the interface mainly experienced changes in the charge density, whereas the changes were more obvious in the S termination of MoSSe. The strongest interaction occurred between the nearest O and S atoms where the charge redistribution is significant. The results are comparable with the study conducted by Y. Li \textit{et al.} although different exchange-correlation functional were used in the calculations \cite{TiO2/MoS2}. In the \ch{TiO2}/MoSeS, notable charge re-arrangement occurred around the interface, which also extended to the outer side of both monolayers. The blue isosurface at the S and Se terminations showed that MoSSe contributes electrons to \ch{TiO2}. In order to quantify the amount of the charge transfer in the heterostructures, we performed the Bader analysis \cite{Bader}. According to the analysis, a charge of 0.038\ \textit{e} and 0.020 \textit{e} per unit cell was transferred from MoSSe to \ch{TiO2} in the \ch{TiO2}/MoSSe and \ch{TiO2}/MoSeS, respectively. Thus, after constructing the heterostructures, n-type doping is realized in \ch{TiO2} while p-type doping in MoSSe. %The electron depletion from MoSSe shifted the VB upward and electron accumulation to \ch{TiO2} induced a CB downshift, leading to the type \rom{2} and type \rom{3} band alignment in the investigated heterostructures. 
The built-in electric field pointing from MoSSe to \ch{TiO2} facilitates the charge separation and suppresses the recombination rate of charge carriers in the \ch{TiO2}/MoSSe \cite{EF1}. Moreover, the larger charge transfer can be attributed to the strong interlayer coupling and narrower interlayer distance, contributing to the larger amount of charge transferred from \ch{TiO2} to MoSSe \cite{C1, C2}. In the \ch{TiO2}/MoSeS electrons were transferred in the opposite direction of the built-in electric field, which is proposed to contribute to reduce charge transfer across the interface. The Bader charges of the individual atoms in the isolated monolayers and heterostructures are provided in Figure S9 and S10, confirming the charge redistribution after constructing the heterostructures. The values support the strongest interaction between the closest O and S(Se) atoms at the interface. The small charge transfer across the interface indicates relatively less chemical interaction between the \ch{TiO2} and MoSSe monolayers. In all, these results indicate that it is possible to construct stable heterostructures, out of lattice matching semiconductor monolayers and tune the electronic properties, according to varying interface terminations ranging from semiconducting to metallic. 

%\section*{Conclusions}
\section{Conclusions}
We have investigated the structural and electronic properties of the \ch{TiO2}/MoSSe and \ch{TiO2}/MoSeS vdW heterostructures using first principles calculations. Both heterostructures were energetically stable, indicated by their negative formation energies. The band alignment was found to depend on the interface termination of MoSSe. The S termination at the interface led to a type \rom{2} band alignment, providing efficient separation of photogenerated electrons and holes. The Se termination at the interface resulted in a type \rom{3} band alignment and enabled a band-to-band tunneling of electrons across the interface. After forming the heterostructures electron transfer occurred from MoSSe to \ch{TiO2}. Interestingly, a built-in electric field was developed inside the heterostructures due to the difference in the work functions of the \ch{TiO2} and MoSSe layers, and that influenced the charge transfer at the interface. Our work demonstrates that the interface termination of MoSSe is a key factor in determining the properties of the \ch{TiO2}-based vdW heterostructure. Tunability via changing the interface termination makes the heterostructure of 2D \ch{TiO2} and MoSSe a potential candidate for various applications.

\section{Acknowledgement}
This work has received funding from the European Union’s Horizon Europe research and innovation program under the Marie Sk\l odowska-Curie grant agreement no. 101081280, and the Finnish cultural foundation (Grant No. 00230235). CSC - IT Center for Science, Finland, is acknowledged for computational resources.

%\section*{Additional content}
%upporting Information Available: 

\section{Declaration of competing interest}
The authors declare no competing financial interests.

\section{Supporting information}
The supplementary material is hosted with the main article.\\
Description: Crystal structure of \ch{MoSSe} and \ch{TiO2} monolayers, different stacking configurations of \ch{TiO2} and MoSSe monolayers, bond lengths in \ch{TiO2}, band structure of \ch{MoSSe} and \ch{TiO2}, partial charge densities and partial density of states of \ch{TiO2}/MoSSe and \ch{TiO2}/MoSeS, planar-averaged electrostatic potential of MoSSe and \ch{TiO2}, and Bader charges in MoSSe and \ch{TiO2}, and in \ch{TiO2}/MoSSe and \ch{TiO2}/MoSSe.

\clearpage

%\section*{For Table of Contents use only}
%\begin{figure}[h!]\centering
%\includegraphics[width=0.52\linewidth]{TOC.jpg}
%\end{figure}

\end{document}

\maketitle
\beginsupplement

\vspace{1cm}

\begin{figure}[h!]\centering
\includegraphics[width=0.4\linewidth]{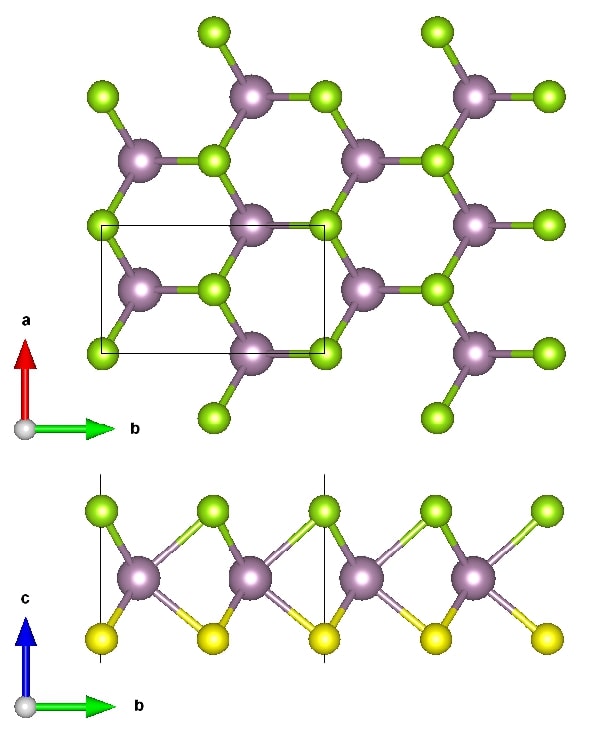}
\caption{Rectangular unit cell of 2D MoSSe with a unit cell size of $a=3.25$\ \AA \ and $b=5.64$\ \AA.}
\label{fig:MLs}
\end{figure}

\begin{figure}[h!]\centering
\includegraphics[width=0.8\linewidth]{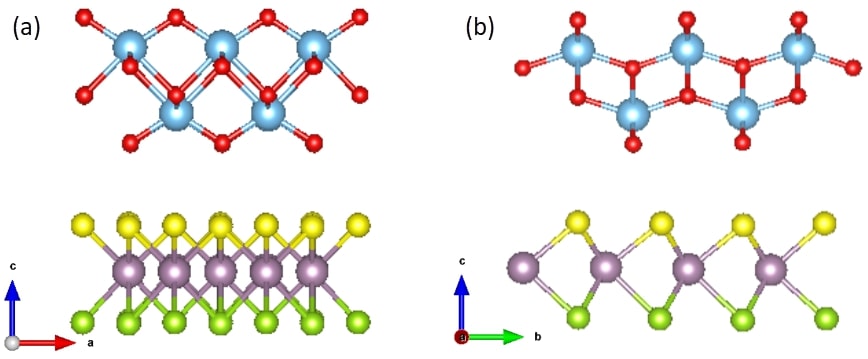}
\caption{Stacking configuration of 2D lepidocrocite-type \ch{TiO2} and 2D MoSSe monolayers along the a) x-direction and b) y-direction. According to Li \textit{et al.} the particular stacking configuration of 2D \ch{TiO2} and \ch{MoS2} is the most stable \cite{TiO2/MoS2}.}
\label{fig:MLs}
\end{figure}

\begin{figure}[h!]\centering
\includegraphics[width=0.8\linewidth]{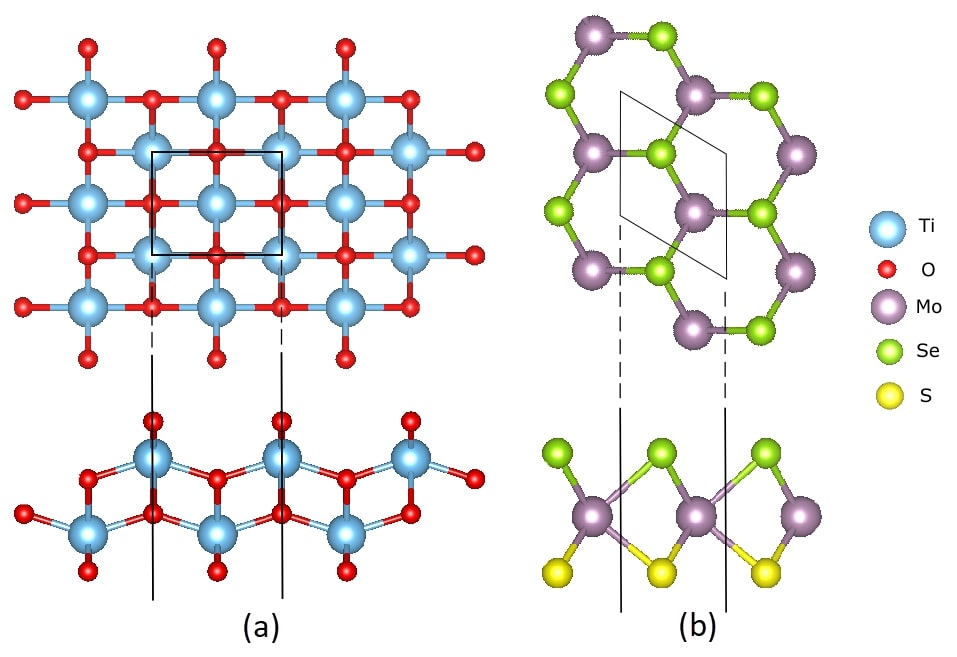}
\caption{Top (upper) and side (lower) view of the a) 2D lepidocrocite-type \ch{TiO2} and b) 2D MoSSe. The primitive unit cells are indicated with black lines.}
\label{fig:MLs}
\end{figure}

\begin{figure}[h!]\centering
\includegraphics[width=0.85\linewidth]{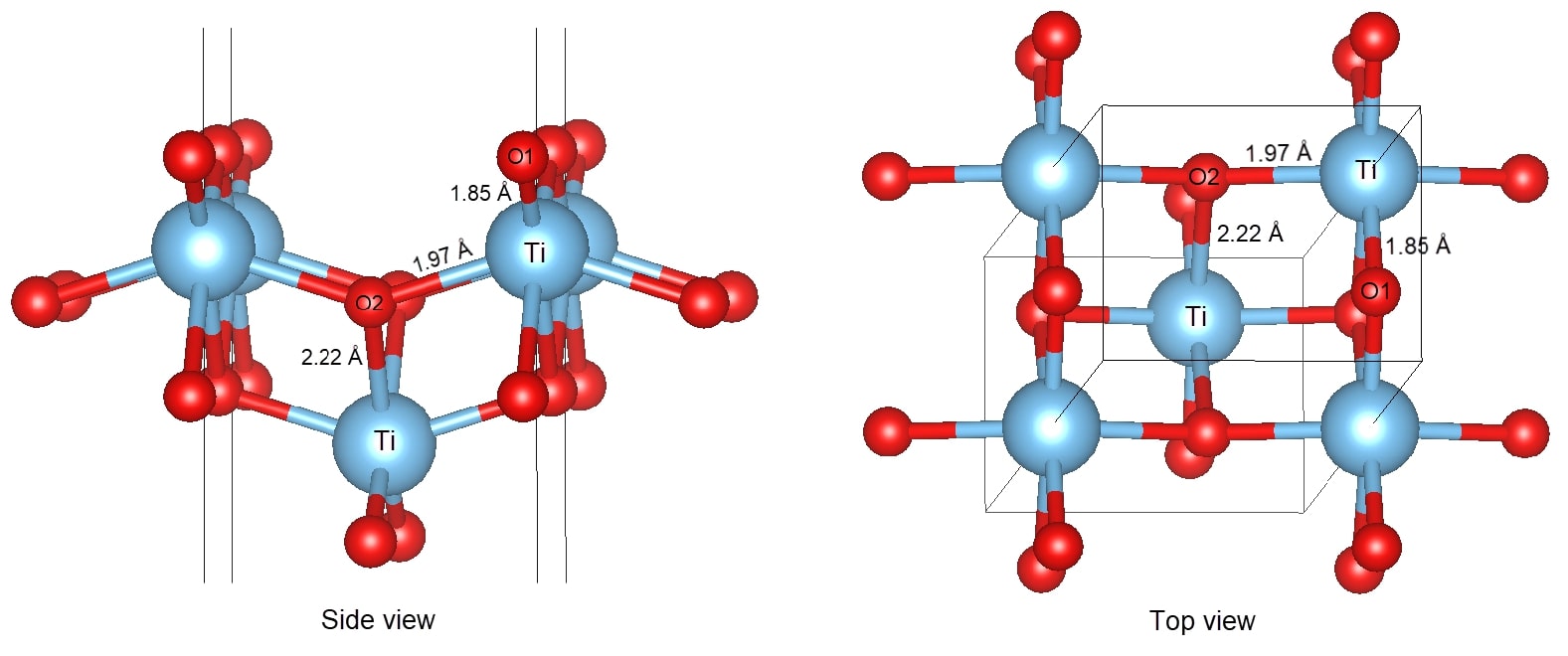}
\caption{The optimized structure of 2D lepidocrocite-type \ch{TiO2}. The monolayer consists of two-fold (O1) and four-fold (O2) oxygen atoms, and six-fold titanium (Ti) atoms. The Ti-O1 bond lengths were found to be 1.85\ \AA, and the two different Ti-O2 bonds were 1.97\ \AA\ and 2.22\ \AA, being in agreement with reported values \cite{bonds1,bonds2}.}
\label{fig:Bonds-TiO2}
\end{figure}

\begin{figure}[h!]\centering
\includegraphics[width=1\linewidth]{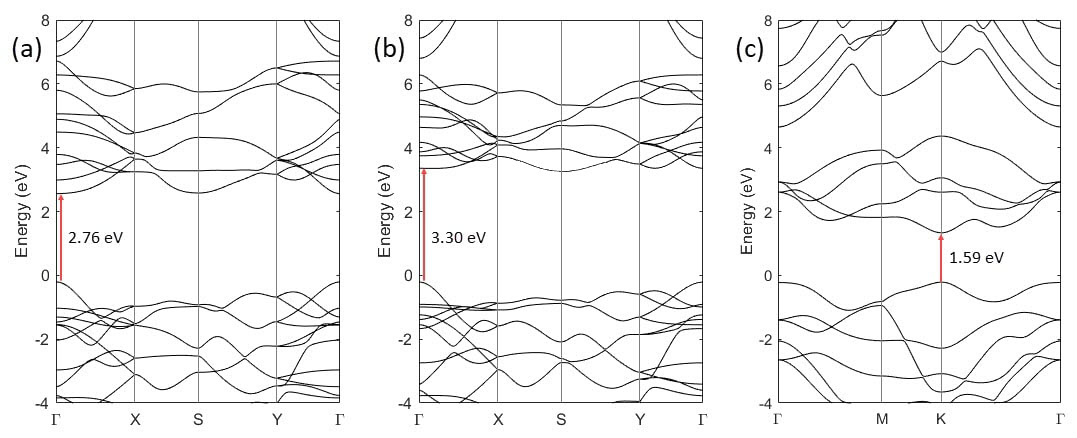}
\caption{Band structure of 2D \ch{TiO2} using the a) GGA and b) GGA+U functional. A direct band gap of 2.76\ eV and 3.30 eV were found, respectively. For c) MoSSe we found a direct band gap of 1.59 eV using the GGA functional. Band gap is indicated with a red arrow in the plots.}
\label{fig:BS-MLs}
\end{figure}

%\begin{figure}[h!]\centering
%\includegraphics[width=0.5\linewidth]{BandStrucTiO2-GGA.jpg}
%\caption{Band structure of 2D lepidocrocite-type \ch{TiO2} using the GGA functional. A direct band gap of 2.76\ eV was found, being in line with the previous theoretical results \cite{LNS-bg1, LNS-bg2}}
%\label{fig:BS-TiO2}
%\end{figure}
\begin{figure}[h!]\centering
\includegraphics[width=1\linewidth]{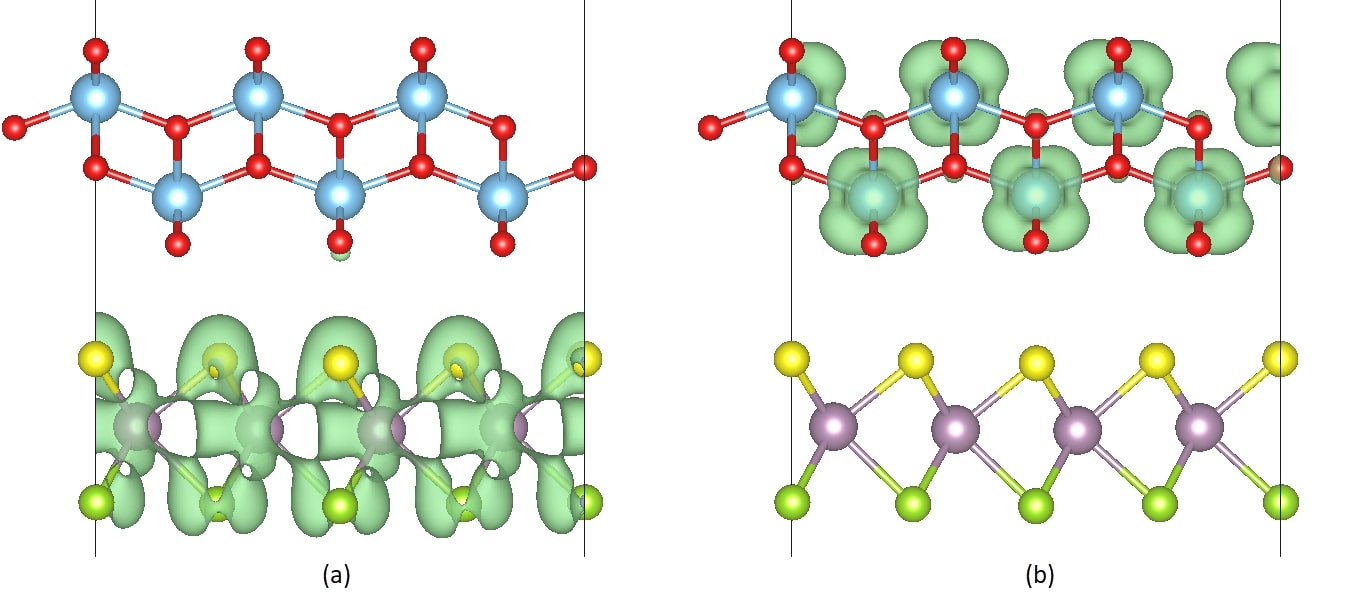}
\includegraphics[width=1\linewidth]{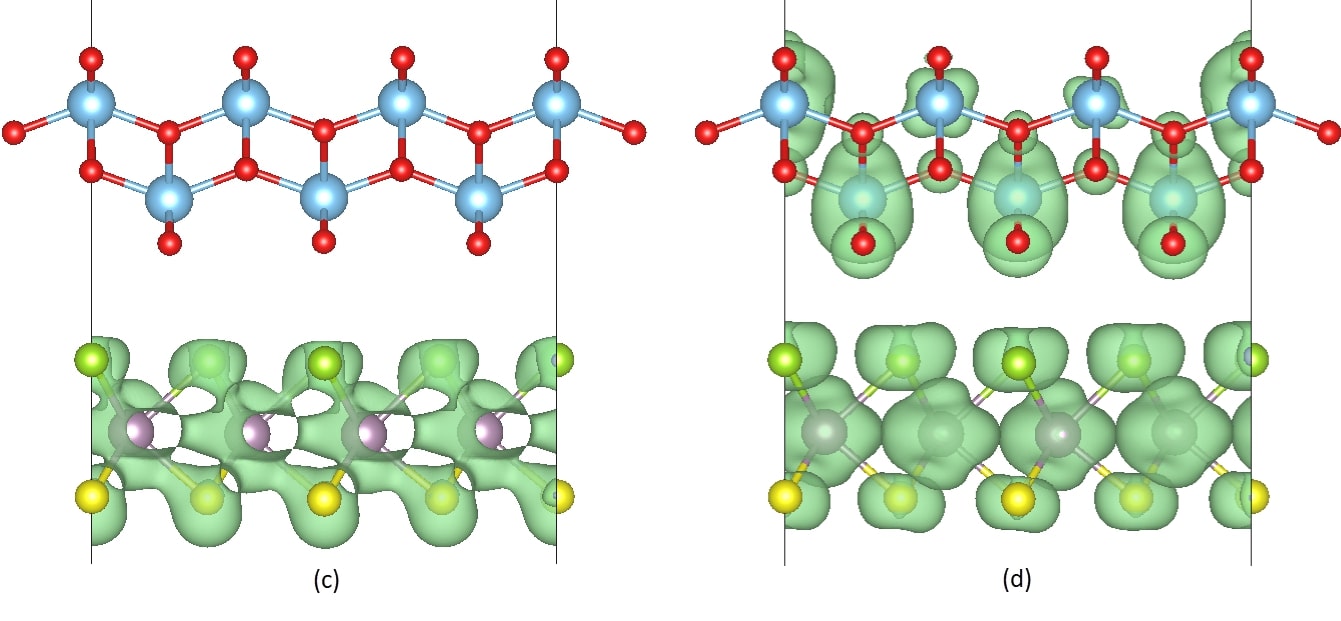}
\caption{Decomposed charge densities of the VBM and CBM of the \ch{TiO2}/MoSSe (a and b) and \ch{TiO2}/MoSeS (c and d). Isosurface value is set to 0.006 \textit{e}\AA$^{-3}$.}
\label{fig:PARCHG}
\end{figure}

\begin{figure}[h!]\centering
\includegraphics[width=0.95\linewidth]{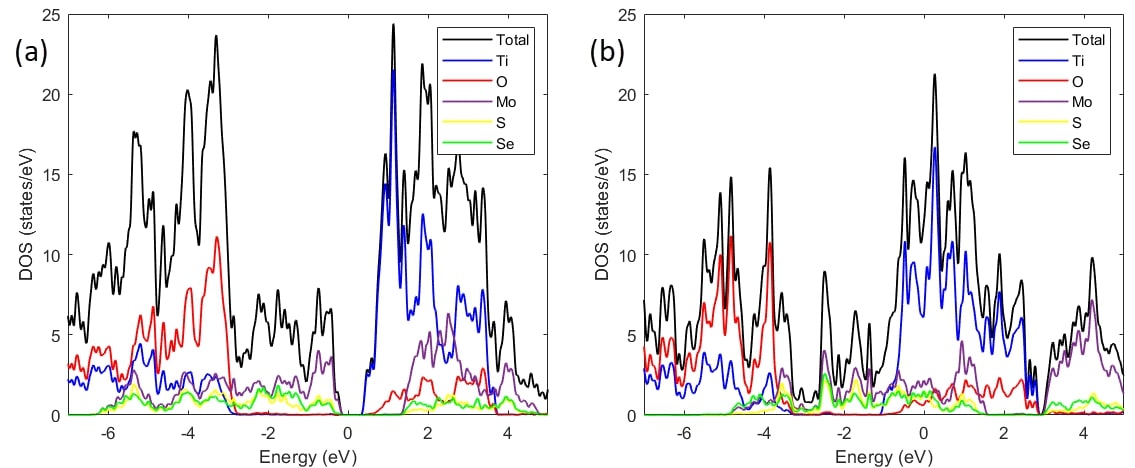}
\caption{Partial density of states of Ti, O, Mo, S and Se in the a) \ch{TiO2}/MoSSe and b) \ch{TiO2}/MoSeS heterostructures.}
\label{fig:PDOS}
\end{figure}

\begin{figure}[h!]\centering
\includegraphics[width=0.75\linewidth]{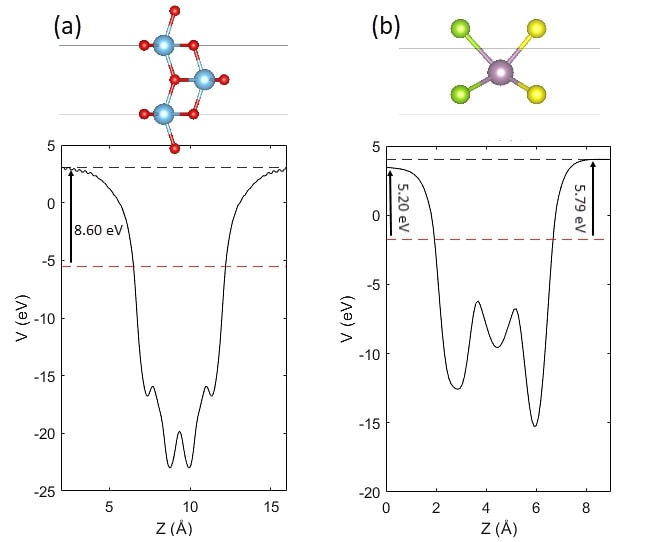}
\caption{Planar-averaged electrostatic potential of the a) \ch{TiO2} and b) MoSSe monolayers.}
\label{fig:PDOS}
\end{figure}

\vspace{0.5cm}

\begin{figure}[h!]\centering
\includegraphics[width=0.75\linewidth]{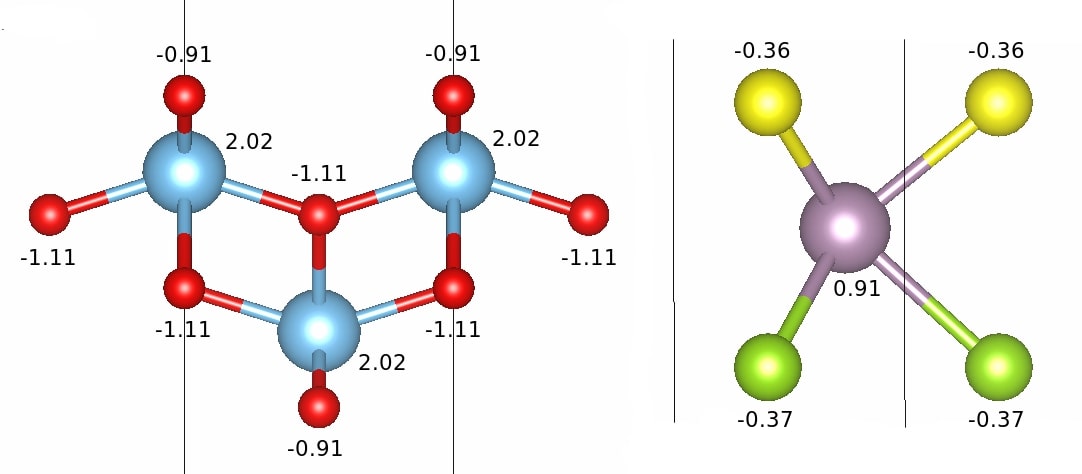}
\caption{Bader charges of the atoms in the free-standing a) \ch{TiO2} and b) MoSSe monolayers. Positive value refers to electron loss and negative value to electron gain. In the \ch{TiO2} Ti atoms give 2.02 \textit{e} for covalent bonding while two-fold and four-fold oxygen atoms gain -0.91 \textit{e} and -1.11 \textit{e} per unit cell, respectively. In the MoSSe Mo atom exhibits a loss of charge which is accumulated to S and Se sites.}
\label{fig:Bader-ML}
\end{figure}

\begin{figure}[h!]\centering
\includegraphics[width=1\linewidth]{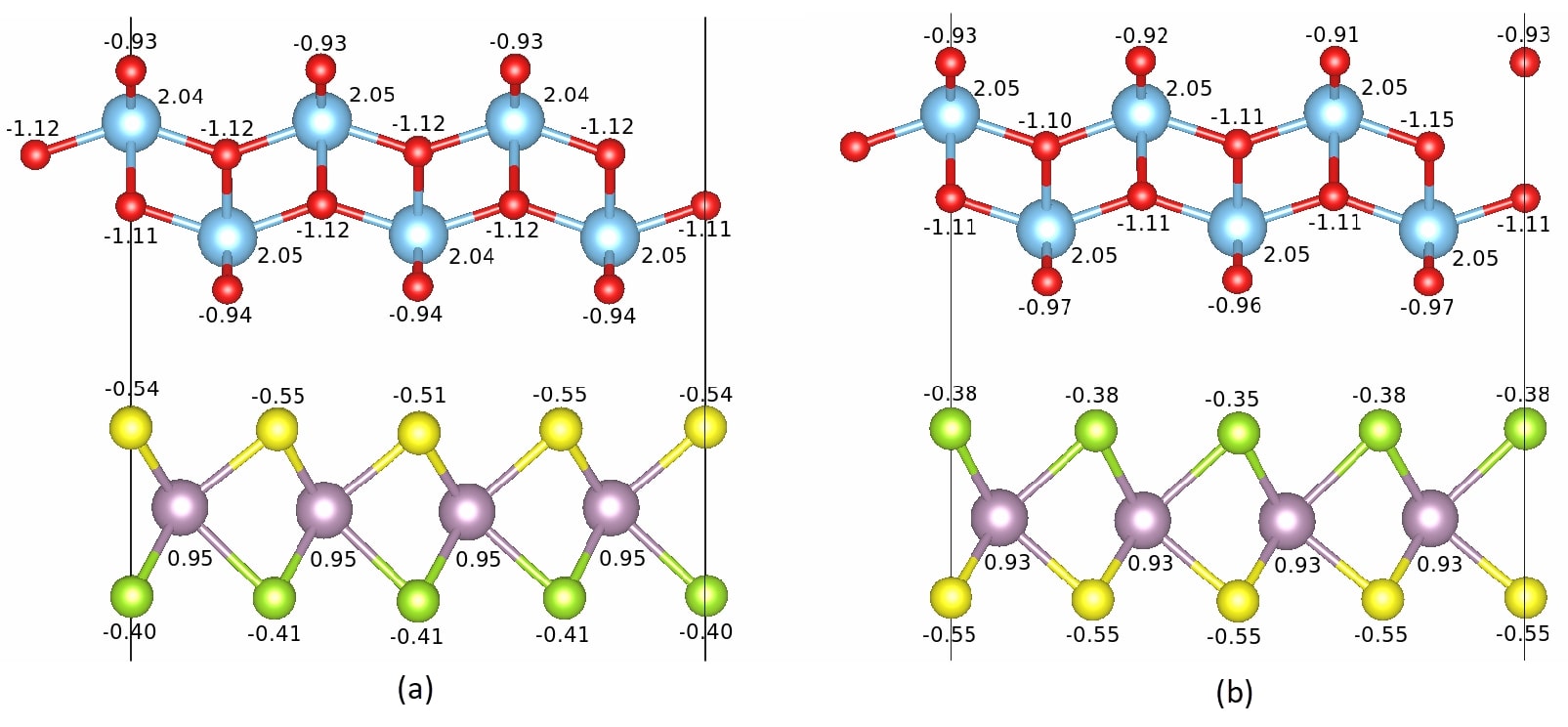}
\caption{Bader charges of the atoms in the a) \ch{TiO2}/MoSSe and b) \ch{TiO2}/MoSeS after constructing the heterostructures. The charge redistribution occurs in both monolayers. At the interface the Bader charges of O atoms and S (Se) atoms vary, showing the strongest interaction between the closest O and S (Se) atoms.}
\label{fig:Bader-HS}
\end{figure}